\documentclass[secnumarabic,amssymb, nobibnotes, aps, prl,twocolumn,superscriptaddress]{revtex4-1}

\usepackage{bm}
\usepackage{graphicx}

\begin{document}

\title{Direct observation of entropic-stabilisation of BCC crystals near melting}%

\author{Joris Sprakel}
\affiliation{School of Engineering and Applied Sciences, Harvard University, Cambridge,  MA 02138, USA.}
\affiliation{Department of Physics, Harvard University, Cambridge,  MA 02138, USA.}
\affiliation{Physical Chemistry and Soft Matter, Wageningen University, Stippeneng 4, 6708 WE, Wageningen, the Netherlands}

\author{Alessio Zaccone}
\affiliation{Department of Chemical Engineering and Biotechnology, University of Cambridge, New Museums Site, Pembroke Street, CB2 3RA Cambridge, UK.}

\author{Frans Spaepen}
\affiliation{School of Engineering and Applied Sciences, Harvard University, Cambridge,  MA 02138, USA.}

\author{Peter Schall}
\affiliation{Van der Waals-Zeeman Institute, University of Amsterdam, Science Park 904, 1098 XH Amsterdam, the Netherlands}

\author{David A. Weitz}
\affiliation{School of Engineering and Applied Sciences, Harvard University, Cambridge,  MA 02138, USA.}
\affiliation{Department of Physics, Harvard University, Cambridge,  MA 02138, USA.}

\email[Corresponding Author: ]{weitz@seas.harvard.edu}

\begin{abstract}
Crystals with low latent heat are predicted to melt from an entropically-stabilised body-centered cubic (BCC) symmetry. At this weakly first-order transition, strongly correlated fluctuations are expected to emerge, which could change the nature of the transition. Here we show how large fluctuations stabilise BCC crystals formed from charged colloids, giving rise to strongly power-law correlated heterogeneous dynamics. Moreover, we find that significant non-affine particle displacements lead to a vanishing of the non-affine shear modulus at the transition. We interpret these observations by reformulating the Born-Huang theory to account for non-affinity; illustrating a scenario of  ordered solids reaching a state where classical lattice dynamics fail.  \\
\end{abstract}


\maketitle

Many common metals, including lithium, sodium and most transition metals of group IV and V, transform from a close-packed structure to a body-centered cubic (BCC) phase at high temperatures. These BCC phases are remarkable as they derive their stability, with respect to other lattices, from a large vibrational entropy\cite{haasen1996physical}. Strongly correlated fluctuations of the lattice are predicted to emerge on approach to the solid-liquid boundary, and, for some BCC crystals, the melting transition becomes only weakly first order\cite{PhysRevLett.41.702}. Similar fluctuations are also expected for melting in the absence of interfaces or dislocations where liquid nucleation is normally initiated\cite{alsayed2005premelting}. This is a different form of melting, first described by Born\cite{:/content/aip/journal/jcp/7/8/10.1063/1.1750497}. These correlated fluctuations could change the very nature of the solid itself. However, these have never been experimentally observed. 

Here, we explore BCC crystals formed by charged colloids and experimentally probe the nature of these stabilizing fluctuations, and their behavior upon approach to melting. We show that the crystals are stabilized at low densities by correlated lattice vibrations in direct analogy to their atomic counterparts. We identify “hot” particles that have the largest fluctuations away from their average lattice positions; they exhibit strong spatial correlation, forming extended connected clusters. These clusters exhibit fractal configurations on the lattice, whose size diverges critically as the solid-liquid transition is approached. This gives the phase transition a pronounced second-order character. However, the affine shear modulus remains non-zero right up to the transition, and thus the transition is strictly first-order. We show that these large fluctuations cause the system to respond non-affinely, and the non-affine shear modulus vanishes continuously. These results demonstrate that thermal excitation of the BCC crystal lead to a new state, where the average structure is crystalline but, surprisingly, long-ranged correlated fluctuations govern the behavior. 

We study charged colloids of poly(methyl methacrylate) of $d = $ 2.1 $\mu$m, suspended in a density-matching mixture of decahydronaphtalene and tetrachloroethylene. The addition of 15 mM AOT, acting as a charge-control agent, leads to the charging of the colloidal particles\cite{doi:10.1021/la702432u} with a screening length $1/\kappa = $ 1 $\mu$m. As a result of the strong electrostatic repulsion, these colloids form solids at very low densities\cite{Yethiraj2003, royall2006re, PhysRevE.91.030301, PhysRevE.68.021407}, which we are known as colloidal Wigner crystals \cite{:/content/aip/journal/jcp/76/7/10.1063/1.443417} by analogy to the crystals formed by electrons on a semiconductor surface \cite{PhysRev.46.1002}. The main control parameter in our experiments is the particle volume fraction $\phi$.  We image each particle in a 90x90x15 $\mu$m volume at 1 Hz with confocal fluorescence microscopy. To determine the short-time dynamics, we complement these data with horizontal two-dimensional slices at 28 Hz.

The phase behaviour of this system as a function of $\phi$ is shown in Figure 1. For $\phi>$ 0.20, the charged colloidal particles form face-centered cubic (FCC) crystals (Fig. 1d); as $\phi$ is decreased the solid adopts a BCC symmetry with few defects \cite{royall2006re,Yethiraj2003} (Fig. 1c). At even lower $\phi$ the sample melts into a liquid (Fig.1a). The  melting of crystalline solids from a superheated state is accompanied by complex dynamics in the form of migrating topological defects \cite{Zhang2013}, particle-exhange loops \cite{Wang05102012} or premelting at defects \cite{alsayed2005premelting}. In our experiments however, we study the approach to melting of samples at a fixed value of $\phi$; to the best of our knowledge, the samples are in equilibrium and thus representative of the thermodynamic phase boundaries of the system. We also note that the measurement of absolute volume fractions in these systems is challenging and may have some error; in the analysis below we present our data as a function of relative volume fractions, such that these errors cancel and do not alter the validity of our conclusions. Previous work has shown a quantitative agreement between the experimental phase behavior of this system and that predicted by theory \cite{PhysRevE.91.030301}, further supporting the assumption of equilibrium.

\begin{figure}[t!]
\begin{center}
\includegraphics[width=\linewidth]{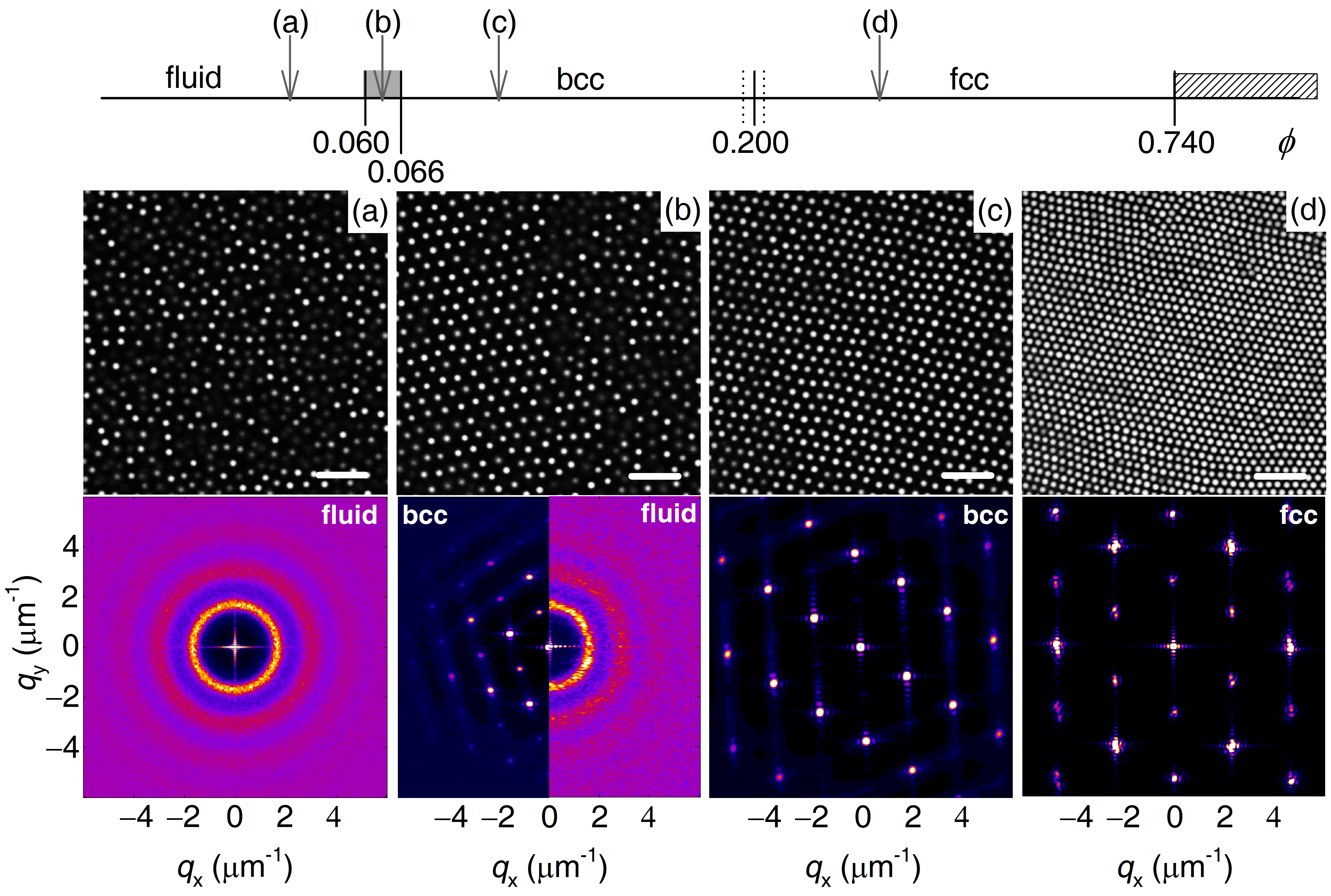}
\caption{\label{fig1} (color online) top) phase diagram as a function of colloid volume fraction $\phi$.  Confocal microscopy images (middle) and structure factors $S(q_x,q_y)$ (bottom) are given for $\phi$= 0.050 (a, fluid), 0.063 (b, fluid-BCC coexistence), 0.130 (c, BCC) and 0.350 (d, FCC); $S(q_x,q_y)$ in b) is separated into that of the coexisting crystal (left) and fluid (right).}
\end{center}
\end{figure}

We observe liquid-crystal coexistence in a narrow range  between 0.060$<\phi<$0.066 (Fig. 1b); this coexistence region is irrefutable evidence of a first-order transition. The long-range order of the crystals is reflected by Bragg peaks in the structure factor (bottom row Fig.1c\&d). By contrast, the liquid sample exhibits only isotropic short-range order (bottom row Fig.1a). For the sample in coexistence, we calculate $S(q_x,q_y)$ for each region separately, and find distinct Bragg peaks for the crystalline region, and isotropic scattering for the liquid region (Fig. 1b).  The  solid-solid transformation from a FCC to a lower density BCC structure which melts into the liquid is in direct analogy to a wide variety of metals, which exhibit a similar transition upon increasing the temperature up to the melting point. Moreover, it is also in accord with the theoretical predictions for crystals with a low latent heat of melting\cite{PhysRevLett.41.702}. The system of charged colloids, which can be analyzed at the single-particle scale, is an ideal model to explore the general scenario of entropically-stabilised BCC phases and their weak first-order transition. 

\begin{figure}[t!]
\begin{center}
\includegraphics[width=\linewidth]{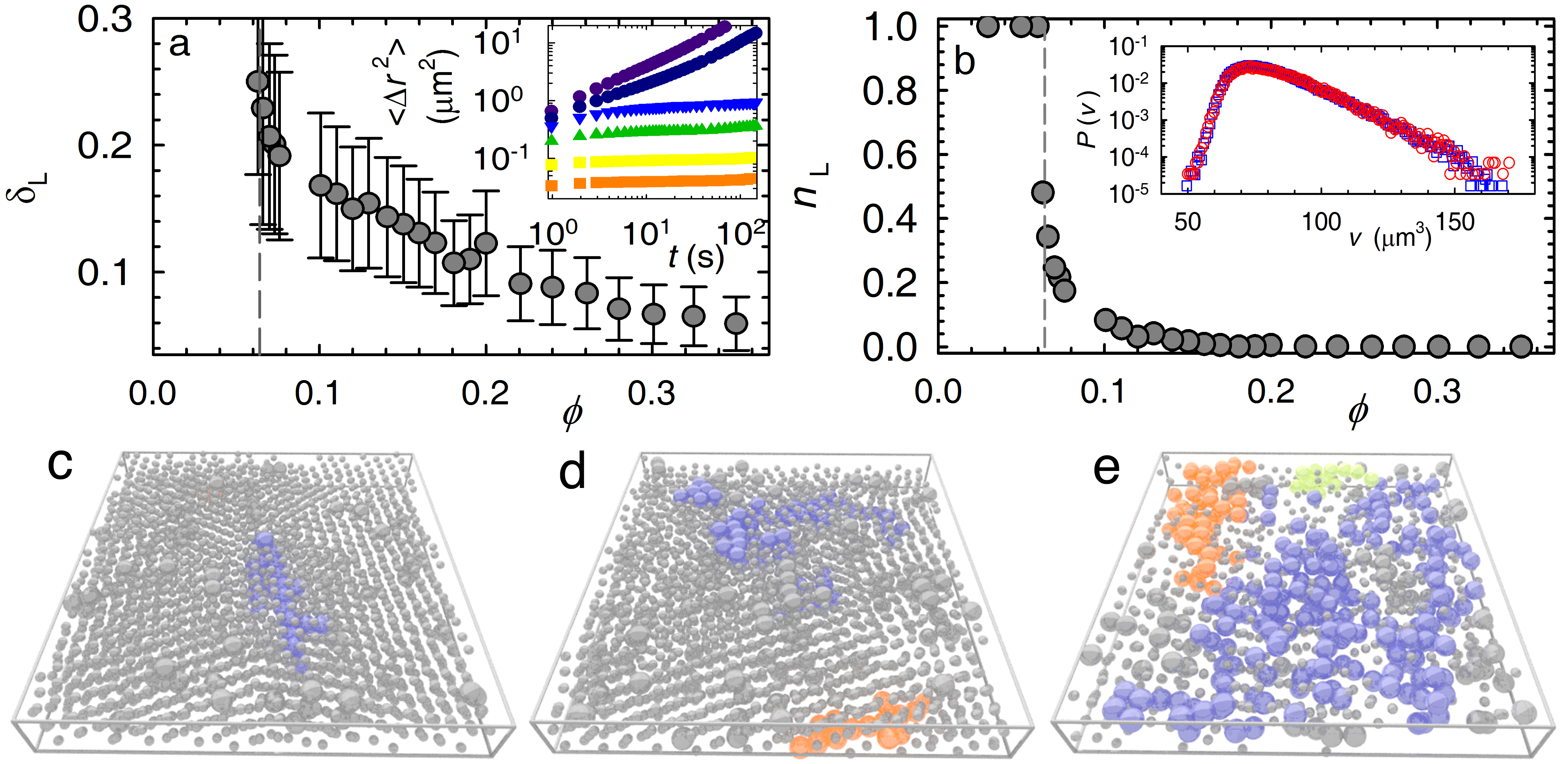}
\caption{\label{fig2} (color online) a) Lindemann parameter $\delta_L$, inset: mean squared displacements $\left< \Delta r^2(t) \right>$ for (top to bottom) $\phi =$ 0.030, 0.060, 0.080, 0.141, 0.221 and 0.301,  b) fraction of "hot" particles $n_L$, inset: distributions of Voronoi volumes $P(v)$ for "hot" particles (red circles) and all others (blue squares), c-e) computer-generated renderings of experimental data, where "hot" particles are color coded according to cluster size, with blue, orange and yellow are 1st, 2nd and 3rd largest clusters respectively, for $\phi-\phi_c = $ 0.095 (c), 0.011 (d) and 0.005 (e).}
\end{center}
\end{figure}

We explore the dynamics of this system by determining the mean-square displacement $\left< \Delta r^2(t) \right>$. We observe two distinct behaviors; a purely diffusive behaviour in the liquid and a time-independent plateau of height $\delta^2$ in the solid (inset Fig. \ref{fig2}a). The time-independent plateau in the solid reflects the amplitude of particle fluctuations around their mean lattice positions. Normalising the amplitude of particle fluctuations with the lattice constant $a$ gives the Lindemann parameter $\delta_L = \delta/a$ . We observe a sharp rise to $\delta_L = $ 0.25 in the solid phase at melting, where the crystal and liquid coexist; beyond this, $\delta_L$ can no longer be defined (Fig.\ref{fig2}a). 

\begin{figure}[t!]
\begin{center}
\includegraphics[width=\linewidth]{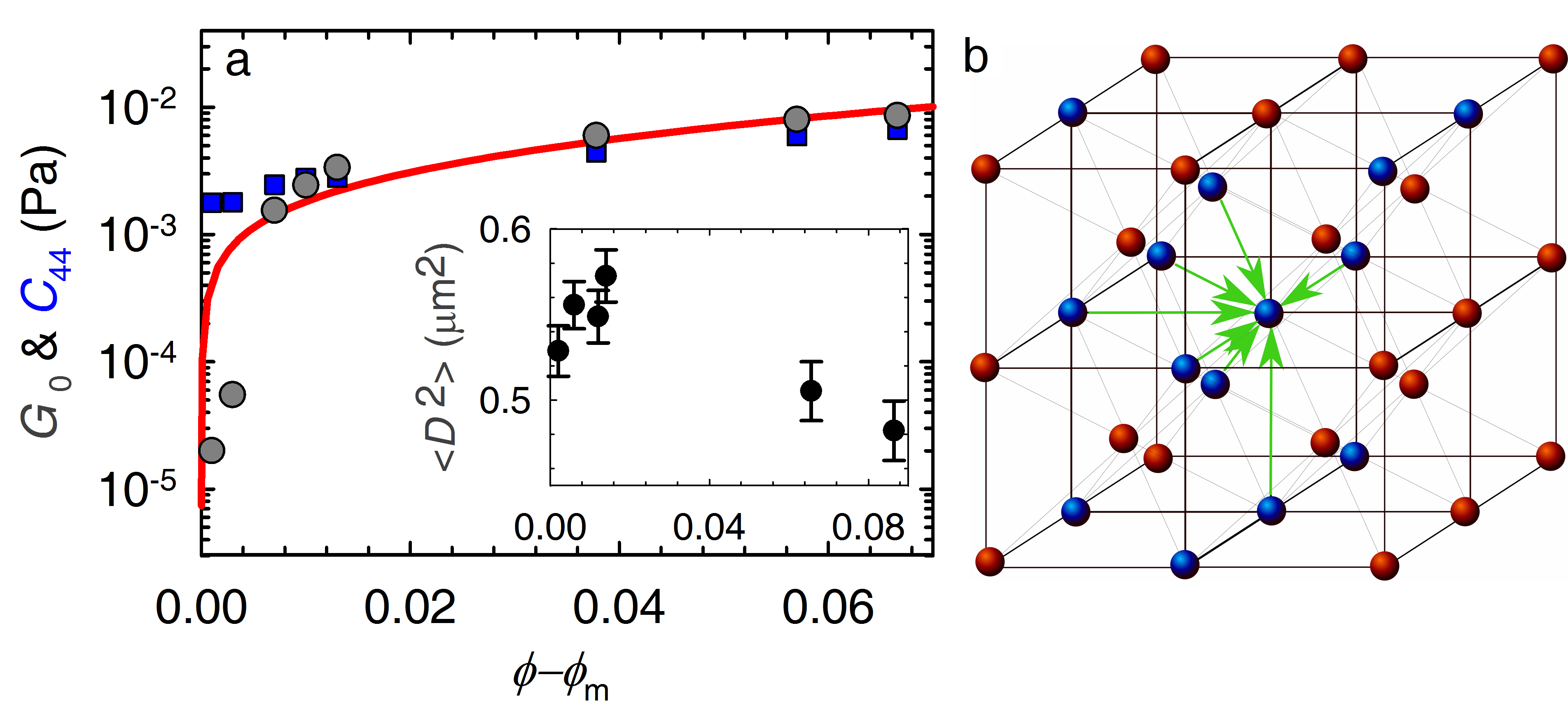}
\caption{\label{fig4} (color online) a) experimental affine elastic constant $C_{44}$ (squares) and nonaffine shear modulus $G_0$ (circles), and prediction from the reformulated Born theory (line). Inset shows the amplitude of non-affine displacements $\left< D^2\right> $, b) illustration of how a heterogeneous distribution of "hot" particles (red) in a BCC lattice disrupts the local force balance between "cold" particles (blue) bonded by green arrows.}
\end{center}
\end{figure}

A defining feature of all solids is a finite shear modulus. The value of the affine shear elastic constant $C_{44}$ is determined by the symmetry of the lattice and the strength of interparticle bonds \cite{:/content/aip/journal/jcp/7/8/10.1063/1.1750497,born1998dynamical}, assuming that all particles are displaced proportionally to the external deformation. From our confocal microscopy data we can directly measure $C_{44}$, as detailed elsewhere\cite{PhysRevE.91.032310}. For volume fractions right up to $\phi_m$, $C_{44}$ remains non-zero (squares in Fig.\ref{fig4}a), decaying as $k_BT/a^3$, consistent with the predictions of affine theory. Once the sample becomes liquid and a lattice can no longer be defined, $C_{44}$ jumps discontinuously to zero. The anisotropy in crystal elasticity for these systems also persists up to the melting transition \cite{PhysRevE.91.032310}. Combined with the distinct Bragg peaks for a crystal coexisting with the liquid (Fig.\ref{fig1}b), these data are in full accord with a transition that is strictly first-order.

Our microscopy data provide a means to investigate the nature of the solid close to this transition, where it exhibits large fluctuations. Indeed, inspection of the images (Fig.\ref{fig1}) and movies  (\emph{see SI}) suggests that large fluctuations create pronounced deformations of the lattice, especially at low $\phi$. To ascertain the nature of these deformations, we identify "hot" particles as those that display instantaneous displacement amplitudes larger than $\delta_L$ = 0.25, the ensemble-averaged value at melting \cite{PhysRevLett.87.055703}. The fraction of "hot" particles in the crystal $n_L$ rises steeply to 0.5 just below the melting transition, whereupon it jumps to a value of 1 in the liquid (Fig. \ref{fig2}b). We find no detectable differences in the local surroundings of "hot" particles and all others (inset Fig.\ref{fig2}b).

\begin{figure}[t!]
\begin{center}
\includegraphics[width=\linewidth]{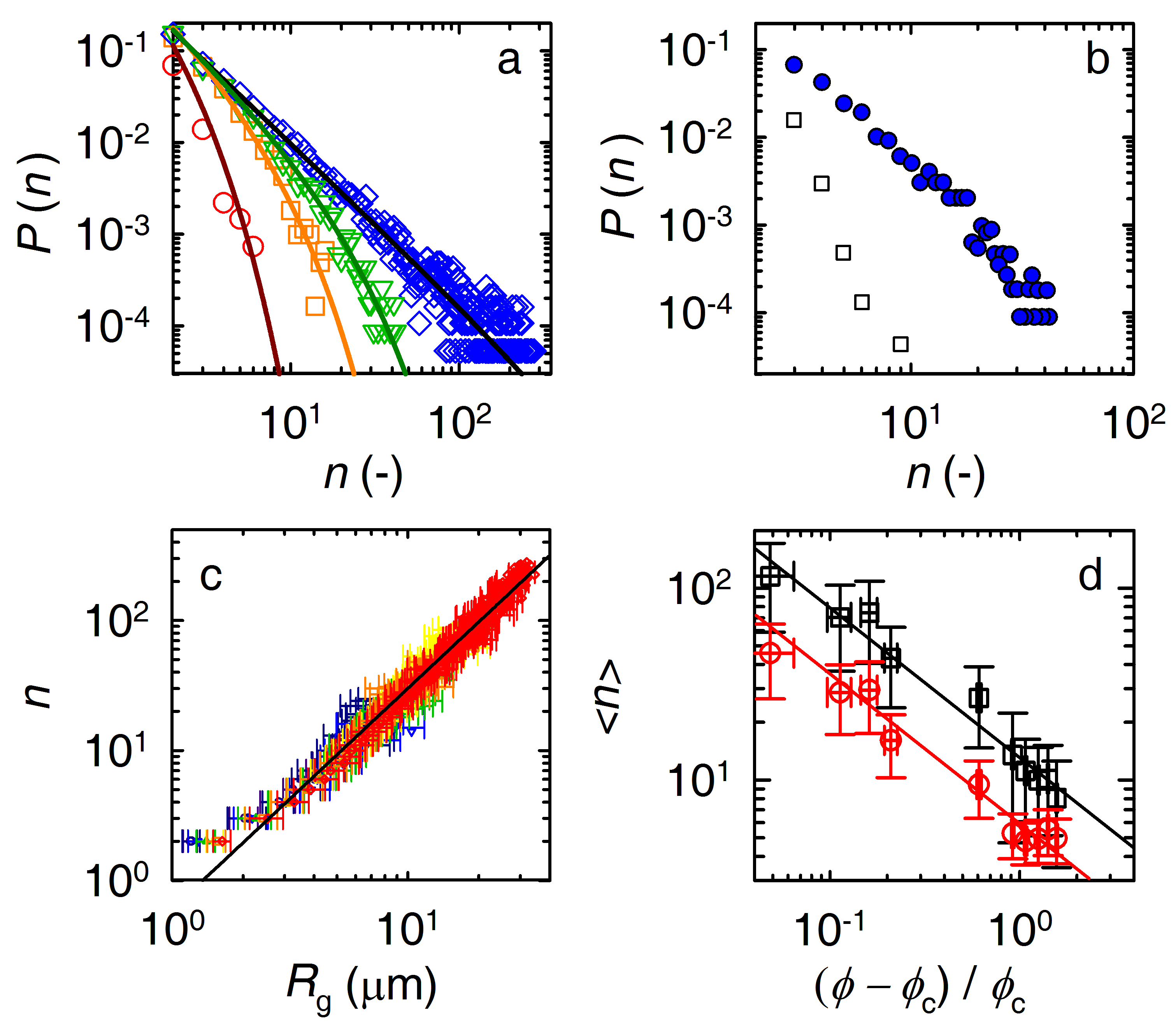}
\caption{\label{fig5} (color online) a) cluster size distributions $P(n)$, for $\phi = $ 0.066 (diamonds), 0.101 (triangles), 0.129 (squares) and 0.190 (circles), drawn lines are fits to a powerlaw distribution with exponential cut-off, b) comparison of experimental $P(n)$ for $\phi = 0.101$ (filled circles) and a distribution for a randomized placements of the same number of hot particles (open squares) c) cluster radius of gyration $R_g$ versus size $n$, for 6 different $\phi$ between 0.17 and 0.066, drawn line indicates $d_f  = 1.70$, , d) characteristic cluster size $ \langle n \rangle$ taken as the time-averaged size of the largest cluster (squares) or the time- and ensemble-averaged size of all clusters (circles), lines indicate a power-law divergence: $\langle n \rangle  \propto \left(\left( \phi - \phi_c \right)/\phi_c \right)^{-3/4}$.}
\end{center}
\end{figure}

The "hot" particles are not homogeneously distributed, but  form connected clusters. We color-code these for several volume fractions in Fig.\ref{fig2}c-e.  Both the size and spatial extent of the clusters increase as the sample approaches the melting transition, where they percolate the field of view (Fig.\ref{fig2}e). These extended and transient clusters are the first observation of the correlated fluctuations implied by the Alexander-McTague theory for any BCC lattice where the difference in free energy between the liquid and solid is small\cite{PhysRevLett.41.702}. We determine their size distribution $P(n)$, with $n$ the number of particles within a connected cluster. We find a distinct power-law correlation with an exponential cut-off (Fig. \ref{fig5}a), whose power-law exponent of 1.75 is independent of volume fraction. We confirm that the clustering of "hot" particles is statistically significant by comparing an experimental $P(n)$ with a simulated distribution for samples in which "hot" particles have been randomly placed on the lattice (Fig.\ref{fig5}b). We also calculate the scaling between cluster size $n$ and their radius of gyration $R_g$ and find that they are fractal in nature, with a characteristic fractal dimension $d_f = $1.7 that is universal for all $\phi$ (Fig.\ref{fig5}c).  

Remarkably, these data exhibit hallmarks of a second-order transition. To demonstrate this we calculate the average cluster size  $\langle n \rangle$; the mean cluster size diverges upon approaching a critical volume fraction $\phi_c$ with a characteristic power-law exponent of $-3/4$ (Fig.\ref{fig5}d). As a confirmation of this critical behavior we also compute the effective cluster volume fraction that also diverges at $\phi_c$ (see SI). Thus, not only do we find heterogeneous dynamics within an on-average perfect crystal, which exhibit fractal correlations in space, but the size and volume fraction of these clusters diverge critically. The critical volume fraction $\phi_c$ for this divergence appears to coincide with the symmetry change at the melting point, which we have shown to be a strictly first-order transition. This provide new insight into the nature of the weak first-order transition described by Alexander and McTague\cite{PhysRevLett.41.702}, but also raises the intriguing question how hallmarks of a continuous transition can coincide with a phase transition which is clearly first-order in nature.

The key to understanding the origin of this behavior lies with the elasticity of the solid. The affine elasticity measured by $C_{44}$ assumes that every particle is displaced exactly proportionally to an applied strain. However, the dynamic disorder caused by the large thermal excitations at low $\phi$ breaks the local BCC symmetry and causes the local coordination number to differ substantially from the value prescribed by a perfect lattice. These transient violations of centre-of-inversion symmetry must result in net non-zero forces on every particle in its affine position (Fig.\ref{fig4}b), which can only be relaxed upon allowing non-affine displacements\cite{PhysRevB.83.184205,PhysRevLett.93.195501}. Such non-affine displacements remove elastic energy from the lattice and thus reduce the overal crystal elasticity. To verify this ansatz, we calculate the amplitude of non-affine displacements\cite{PhysRevLett.107.198303} $\left< D^2 \right>$; we indeed find that $\left< D^2 \right>$ grows upon approaching the melting transition, where the deviations of perfect lattice order are largest (inset \ref{fig4}a). Clearly, any consideration of the crystal rigidity under these conditions must take non-affinity into account. 

We measure the non-affine shear modulus $G_0$  from the correlations in motion between pairs of particles, averaged isotropically and over all pairs in the field-of-view\cite{PhysRevLett.85.888}. Strikingly, the non-affine rigidity $G_0$ vanishes at $\phi_c$ (circles in Fig.\ref{fig4}a), where the affine modulus $C_{44}$ remains non-zero. This confirms the importance of non-affinity due to thermal fluctuations as the crystal approaches melting. We note that the non-affine modes and a vanishing non-affine modulus are crucial features of several approaches to describe disordered solids \cite{0953-8984-22-3-033101}, yet they have remained largely unexplored for crystals to date.

A remarkable paradox emerges from these observations; whereas the phase transition is strictly  first-order, we measure a continuous vanishing of the non-affine shear rigidity and divergence of collective length scales, akin to a second-order transition. The affine elastic constants are based on the symmetry of a perfect lattice; thus $C_{44}$ must be discontinous as the symmetry changes discontinuously at $\phi_m$. By contrast, the non-affine modulus $G_0$ does not require assumption of a specific symmetry and is thus very sensitive to thermal disorder. In effect, the non-affine modulus provides a local probe of a more random random environment with a symmetry much closer to the liquid; this results in a transition which is continuous to within experimental resolution.

For an ordered network of springs deforming affinely, the affine elastic modulus can be estimated as\cite{:/content/aip/journal/jcp/7/8/10.1063/1.1750497,born1998dynamical} $kT/a^3$. By contrast, our experiments show a non-affine modulus which is order-of-magnitude below this limit. Naively, we can set the typical length scale of the clusters of "hot" particles $R_g$ as the relevant scale governing the non-affine mechanics in proximity to $\phi_c$ as $G_0 \propto kT/R_g^3$, which can be related to the cluster statistics as $R_g^3 \propto \langle n \rangle^{3/d_f}$. From our experiments we know that $\langle n \rangle \propto (\phi-\phi_c)^{-3/4}$; thus $G_0 \propto (\phi - \phi_c)^{9/4d_f}$. Remarkably, this simple argument explains two key experimental observations: the vanishing of the non-affine modulus at $\phi_c$ and the convergence of $G_0$ and the affine $C_{44}$ at high volume fractions where $R_g \approx a$.

To quantitatively explain our results, we must explicitly account for the non-affinity by extending the original Born-Huang theory\cite{:/content/aip/journal/jcp/7/8/10.1063/1.1750497,born1998dynamical}. Non-affine displacements lower the free energy of deformation and hence reduce the nonaffine shear modulus $G_0$ with respect to its affine counterpart $C_{44}$: $G_0=C_{44}-G_{NA}$, where $G_{NA}$ is the nonaffine correction\cite{PhysRevLett.110.178002,PhysRevB.83.184205}. To find $G_{NA}$, we adopt a framework for the rigidity of networks with central force bonds. The mechanical stability is governed by the distance from isostaticity $Z - Z_c$, where $Z$ is the average number of bonds at each node and its critical value $Z_c = $6 defines the isostatic point\cite{PhysRevLett.110.178002,PhysRevB.83.184205}. The coordination number $Z$ represents the number of stress-bearing, permanent, bonds with neighbors; in this case these are the bonds not part of the clusters of "hot" particles. The stability parameter $Z-Z_c$ can be directly related to experimentally measurable properties of the clusters, such as $d_f$ and their pair correlation function.\cite{Ma1985} Within this approach, the full derivation of which is given in the SI, we predict the shear modulus to vanish as $G_0=G_A-G_{NA}=K \phi \left(\phi-\phi_c \right)^{0.56}/a \propto \phi\left(\phi - \phi_c\right)^{0.64}$, in which $a$ is the lattice spacing and $K$ a proportionality constant, which is the only adjustable parameter in our model. Remarkably, this theoretical prediction is in quantitative agreement with the experiments as shown by  the solid line in Figure \ref{fig4}a. 

These results give rise to an unexpected picture of the behaviour of entropically-stabilised BCC crystals, which by nature are subject to strong thermal excitations. Due to the inherently low coordination number in the BCC phase, softening of the crystal triggers the emergence of strongly correlated heterogeneous dynamics on the lattice, while the average structure of the crystal lattice remains perfect. The correlated fluctuations and associated non-affine mechanics exhibit the hallmarks of a continuous, critical, transition, that paradoxically coincides with the strictly first-order melting point of the crystal. Such large collective fluctuations increase the entropy of the solid\cite{Frenkel2015}, which extend the crystal stability to lower densities and lead to a very small jump in enthalpy at the first-order solid-liquid transition. They are also observed for various atomic crystals such as those formed by sodium or lithium. Moreover, these correlated fluctuations provide the mechanism for the elastic collapse that causes melting of a superheated crystal that was first anticipated by Born.  

Our observation of strongly correlated fluctuations are unique to the BCC phase, in contrast to, for example, the more common FCC structure in colloidal crystals. More than other structures, the BCC structure is stabilized by entropy with respect to the liquid \cite{haasen1996physical}. As a result, its first-order melting transition can become sufficiently weak that the effects of non-affinity become significant. We also find the high density FCC phase in our experiments (Fig.\ref{fig1}), which is stable down to $\phi = 0.2$. Even at these low densities, and at $\delta_L \approx 0.1$, where the hard-sphere FCC would melt \cite{alsayed2005premelting}, the FCC crystals do not show any non-affinity. This is corroborated by experiments on FCC crystals formed from colloidal hard spheres, in which no deviations from continuum lattice dynamics were observed \cite{ghosh2011low}. Finally, we notice that the affine and non-affine moduli converge upon approaching the FCC phase (Fig.\ref{fig4}a). Non-affine displacements are therefor not of significance for the FCC symmetry, but that they are a particular feature of the high-temperature, or low density, BCC phase. This is in full agreement with the predictions of Alexander-McTague\cite{PhysRevLett.41.702}.

The collective fluctuations we observe increase the entropy of the solid\cite{Frenkel2015} and lead to a very small jump in enthalpy at the first-order solid-liquid transition, also observed for various atomic crystals such as those formed by sodium or lithium. Our results for colloidal crystals may thus help in understanding  weak BCC phases near melting in a much wider variety of systems. For example, over 40 elements in the periodic table exhibit a high-temperature BCC phase close to their melting line. Moreover, low-density BCC crystals of charged particles are  of interest in astrophysics, as they are expected to be an important state of matter in neutron stars and pulsars \cite{baym1971neutron, engstrom2016crystal}. Our results illustrate a scenario in which large thermal fluctuations may bring perfectly ordered solids to a state where classical theories for lattice mechanics break down, and new, richer physics emerges. \\

\section*{Acknowledgements} This work was supported by the National Science Foundation (DMR-1310266, DMR-1206765), the Harvard Materials Research Science and Engineering Center (DMR-1420570) and NASA (NNX13AQ48G). The authors thank Peter J. Lu and Emily Russell for data analysis routines.


%

\end{document}